\documentclass[runningheads]{llncs}
\usepackage{graphicx}
\usepackage{url}
\usepackage{tcolorbox}
\usepackage{caption}
\usepackage{subcaption}
\usepackage{csquotes}
\usepackage{algorithmic}
\usepackage{amssymb}
\usepackage{amstext}
\usepackage{amsmath}
\usepackage{multicol}
\usepackage{hyperref}
\usepackage{multirow}
\usepackage{algorithm}
\usepackage{booktabs} % For formal tables
\usepackage[framemethod=tikz]{mdframed}

\mdtheorem[style=theoremstyle]{definition}{Definition}

\begin{document}
%
%\linenumbers
\title{XtraLibD: Detecting Irrelevant Third-Party libraries in Java and Python Applications}
\titlerunning{XtraLibD: Detecting Irrelevant Third-Party Libraries}
\author{Ritu Kapur\inst{1}\orcidID{0000-0001-7112-0630} \and
Poojith U Rao\inst{1} \and
Agrim Dewam\inst{2} \and
Balwinder Sodhi\inst{1}
}
\authorrunning{R. Kapur et al.}
\institute{Indian Institute of Technology, Ropar, Punjab  , India\\ 
\email{dev.ritu.kapur@gmail.com,\{poojith.19csz0006,sodhi\}@iitrpr.ac.in}\\
 \and
Punjab Engineering College, Chandigarh, India.\\
\email{agrim334@gmail.com}
}
\maketitle              % typeset the header of the contribution
\begin{abstract}
Software development comprises the use of multiple Third-Party Libraries (TPLs). However, the irrelevant libraries present in software application's distributable often lead to excessive consumption of resources such as CPU cycles, memory, and modile-devices' battery usage. Therefore, the identification and removal of unused TPLs present in an application are desirable. We present a rapid, storage-efficient, obfuscation-resilient method to detect the irrelevant-TPLs in Java and Python applications. Our approach's novel aspects are \textit{i)} Computing a vector representation of a .class file using a model that we call Lib2Vec. The Lib2Vec model is trained using the Paragraph Vector Algorithm. \textit{ii)} Before using it for training the Lib2Vec models,  a .class file is converted to a \textit{normalized} form via semantics-preserving transformations. \textit{iii)} A  e\textit{Xtra} \textit{Lib}rary \textit{D}etector (XtraLibD) developed and tested with 27 different language-specific Lib2Vec models. These models were trained using different parameters and $>$30,000 .class and $>$478,000 .py files taken from $>$100 different Java libraries and 43,711 Python available at MavenCentral.com and Pypi.com, respectively.  XtraLibD achieves an accuracy of 99.48\% with an F1 score of 0.968 and outperforms the existing tools, viz., LibScout, LiteRadar, and LibD with an accuracy improvement of 74.5\%, 30.33\%, and 14.1\%, respectively. Compared with LibD, XtraLibD achieves a response time improvement of 61.37\% and a storage reduction of 87.93\% (99.85\% over JIngredient). Our program artifacts are available at \texttt{\url{https://www.doi.org/10.5281/zenodo.5179747}}.

\end{abstract}

\section{Introduction}
Third Party Libraries (TPLs) play a significant role in software development as they provide ready-made implementations of various functionalities, such as image manipulation, data access and transformation, and advertisement. As reported by \cite{liu2015efficient},  57\% of apps contain third-party ad libraries, and 60\% of an application's code is contributed by TPLs \cite{wang2015wukong}. However, as the software development process progresses through multiple iterations, there is generally a change in the requirements or technology. In the process of performing modifications to embed the changes into the application, an unused (or irrelevant) set of libraries (which were used earlier) often remains referenced in the application's distributable. Such unused TPLs have become a prominent source of software bloat. We refer to such unused TPLs as irrelevant-TPLs. Resource wastage is a critical problem for mobile devices that possess limited computing resources and significantly impact the performance by affecting the execution time, throughput, and scalability of various applications \cite{mitchell2007causes,xu2010software}.   
Therefore, the \textbf{identification and removal of the irrelevant TPLs} present in an application are desirable.

\subsection{Motivation}
\label{sec:motivation}
Our primary objective is to develop a technique for detecting \emph{irrelevant TPLs}  present in a software application's distributable binary. An essential task for achieving this objective is to \emph{develop a robust technique for TPL-detection}. The existing TPL-detection methods \cite{backes2016reliable,ma2016libradar} generally depend, directly or indirectly, on the package names or the library's structural details and code. Thus, they are potentially affected by various obfuscations, such as package-name obfuscation and API obfuscation. Also, most of the works are restricted to Android applications \cite{zhang2018detecting,ma2016libradar,li2017libd,backes2016reliable}. 

\begin{tcolorbox}[boxrule=1pt]
\begin{definition}[Irrelevant-TPL]
\label{def:irrelevant-libraries}
We define an \textbf{irrelevant TPL} as the one bundled in the distributable binary of a software application $A$ but not relevant to it. The examples of such TPLs would be the Mockito\footnote{https://site.mockito.org/} or JUnit\footnote{https://junit.org/} Java ARchives (JARs) that get packaged in the deployable release archive of a Java or Python application.

The relevance of a TPL is based on its application in different contexts. For instance, relevant vs. irrelevant, reliable vs. unreliable, anomalous vs. non-anomalous, bloat vs. non-bloat, used vs. unused etc. The idea is to compare with a reference list of relevant libraries or white-listed libraries\footnote{Black libraries matter} in an automated manner.
\end{definition}
\end{tcolorbox}

\begin{tcolorbox}[boxrule=1pt]
\begin{definition}[Paragraph Vector Algorithm]
\label{def:pva}
\textbf{Paragraph Vector Algorithm (PVA)} is an unsupervised algorithm that learns fixed-length feature representations from variable-length pieces of texts, such as sentences, paragraphs, and documents. The algorithm represents each document by a dense vector trained to predict words in the document \cite{le2014distributed}.
\end{definition}
\end{tcolorbox}

In our recent work, we proposed an obfuscation-resilient tool, Bloat Library Detector (BloatLibD) \cite{dewan2021bloatlibd}\footnote{Tool and dataset available at~\url{https://www.doi.org/10.5281/zenodo.5179634}}, that detects the TPLs present in a given Java application by identifying the similarities between the source code of the \enquote{available TPLs} and the TPLs used in the given application. To obtain this set of \enquote{available TPLs,} we leverage the TPLs present at MavenCentral repository\footnote{\url{https://mvnrepository.com/repos/central}} that hosts TPLs for various functionalities, which we assume to be a trustworthy source. In our previous work \cite{dewan2021bloatlibd}, we also validated the efficacy of  PVA in computing a reliable and accurate representation of binary and textual code. The current work aims to extend the existing method for Python applications. We name our new tool as eXtra Library Detector (XtraLibD) which is capable of detecting irrelevant libraries in both Java and Python applications. In our current work, we also compare XtraLibD for some new TPL detection tools for Python and Java applications. One of the reasons for choosing these very languages is their popularity with the professional programmers \cite{so-survey-2020} and the availability of a large volume of OSS source code written using these languages\footnote{Sources of stats: https://octoverse.github.com/projects.html and https://githut.info/}.

\begin{figure*}[ht]
\centering
     \begin{subfigure}[b]{\columnwidth}
         \centering
         \includegraphics[width=\columnwidth]{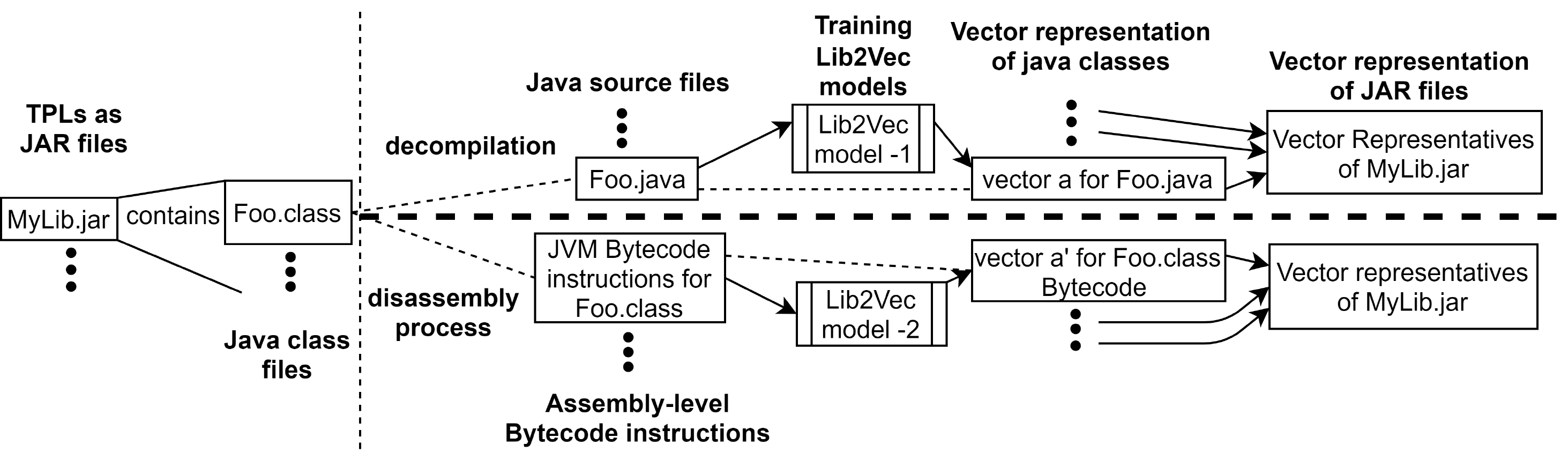}
         \caption{For Java TPLs \cite{dewan2021bloatlibd}}
         \label{fig:javaTPLs}
     \end{subfigure}
     \hfill
     \begin{subfigure}[b]{\columnwidth}
         \centering
         \includegraphics[width=\columnwidth]{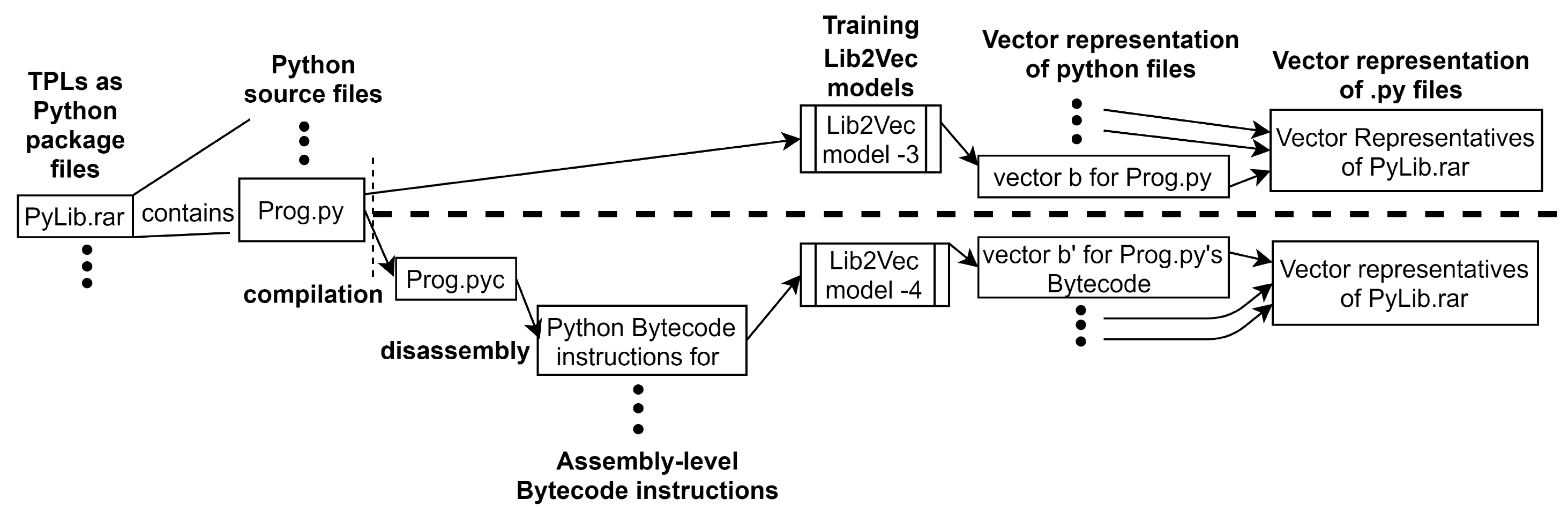}
         \caption{For Python TPLs}
         \label{fig:pythonTPLS}
     \end{subfigure}
    \caption{\centering{Basic idea of our method.}}
    \label{fig:basic-idea}
\end{figure*}

\subsection{Basic tenets behind our system}
\label{sec:basic-idea}

In this paper, we  present a novel TPL-detection technique by creating a \textit{library embedding} using PVA -- we named it Lib2Vec. The central idea underlying our approach is illustrated in Figure~\ref{fig:basic-idea} and stated as follows: 
\begin{enumerate}
    \item Each of the TPLs consists of a collection of binary .class files or source code files, which we refer to as TPL-constituent files.
    
    \item Semantics-preserving transformations (such as compilation, decompilation, and disassembly) are applied to the TPL-constituent files to obtain their \textit{normalized} textual form(s), viz., the textual forms of source code and bytecode instructions.
    
    \item With a large corpus of such text, we train Lib2Vec models, using which a vector representation of any TPL-constituent file can be computed.

   \item Further, the vector representations of a TPL can be computed as a suitable function of the vector representations of all the TPL-constituent files contained in that TPL.
   \item If the vector representations of a TPL $T$ in an application, have a considerable cosine similarity\footnote{\url{http://bit.ly/2ODWoEy}} with the vector representations of the set of \enquote{available TPLs,} we label $T$ as \emph{likely-to-be-non-irrelevant-TPL} or else \emph{likely-to-be-irrelevant-TPL}.
\end{enumerate}

\subsection{Handling obfuscated libraries}
One of the significant issues faced in TPL-detection is the obfuscation of the library code. The TPL-detection techniques that rely on the obfuscation-sensitive features of a library would fail to detect a library under obfuscation. The key idea underlying our approach towards handling obfuscated libraries is to produce a \enquote{normalized} textual representation of the library's binary .class files before using it to train the Lib2Vec models and when checking an input .class using a Lib2Vec model. We perform the decompilation and disassembly of .class files to obtain their \enquote{normalized} textual forms, viz., source code and bytecode instructions as text.
These operations on a .class file are obfuscation-invariant. Similarly, for Python cose, we perform compilation operation on .py files present in .zip package of a Python TPL to obtain .pyc files, from which th bytecode instructions are obtained by performing the dissasembly operation. For example, we transform a .class file using a decompiler \cite{procyon} (with suitable configuration), which produces almost identical output for both obfuscated and unobfuscated versions. The decompiler can emit either bytecode or the Java source code for the .class files.

%\section{Categorization of the existing TPL-detection techniques}
\section{\uppercase{Related Work}}
\label{sec:literature-review-TPL-tools}
Most of the existing works that target TPL-detection assume that \enquote{\emph{the libraries can be identified, either through developer input or inspection of the application's code.}} The existing approaches for TPLs detection can be categorized as follows:
 \begin{enumerate}
     \item \emph{Based on a  \enquote{reference list} of TPLs:} The techniques presented in \cite{chen2014achieving,liu2015efficient,grace2012unsafe,book2013longitudinal} are significant works in this category. A \enquote{reference list} comprises libraries  known to be obtained from a trustworthy source and useful for software development. The basic idea behind the approach is first to construct a \enquote{reference list} of libraries and then test the application under consideration using the list. In this process, it is evaluated that the application's constituent libraries are present in the \enquote{reference list} or not.  All the constituent libraries, which are not present in the list, are deemed to be irrelevant-TPLs. In practice, this approach requires keeping the \enquote{reference list} up-to-date. Since these methods require manually comparing the libraries with the \enquote{reference list} and a periodic update of this list,  they tend to be slower, costly, and  storage-inefficient.
  
     \item \emph{Features-Based approaches:} \cite{backes2016reliable,ma2016libradar,li2017libd} are some of the approaches that work by extracting individual libraries' features and then use them to identify libraries that are similar to another. The feature-based methods generally depend, directly or indirectly, on the package names or the structural details of the application and source code.
     A brief description of these works is provided below:  
     \begin{enumerate}
         \item \textbf{LibScout \cite{backes2016reliable}} presents a TPL-detection technique based on Class Hierarchical Analysis (CHA) and hashing mechanism performed on the application's package names.  Though the method has been proven to be resilient to most code-obfuscation techniques, it fails in certain corner cases. For example, modification in the class hierarchy or package names, or when the boundaries between app and library code become blurred. Another recent work is \cite{feichtner2019obfuscation}, which relies on the obfuscation-resilient features extracted from the Abstract Syntax Tree of code to compute a code fingerprint. The fingerprint is then used to calculate the similarity of two libraries.
         
         \item \textbf{LibRadar \cite{ma2016libradar}} is resilient to the package name obfuscation problem of LibScout and presents a solution for large-scale TPL-detection. LibRadar leverages the benefits of hashing-based representation and multi-level clustering and works by computing the similarity in the hashing-based representation of static semantic features of application packages. LibRadar has successfully found the original package names for an obfuscated library, generating the list of API permissions used by an application by leveraging the API-permission maps generated by PScout \cite{au2012pscout}. Though LibRadar is resilient to package obfuscation, it depends on the package hierarchy's directory structure and requires a library candidate to be a sub-tree in the package hierarchy. Thus, the approach may fail when considering libraries being packaged in their different versions \cite{li2017libd}. An alternate version of LibRadar, which uses an online TPL-database, is named as LiteRadar.  
        
        \item \textbf{LibD \cite{li2017libd}} leverages feature hashing to obtain code features, which reduces the dependency on package information and supplementary information for TPL-detection. LibD works by developing library instances based on the package-structure details extracted using Apktool, such as the direct relations among various constituent packages, classes, methods, and homogeny graphs.  Authors employ Androguard \cite{androguard} to extract information about central relationships, viz., inclusion, inheritance, and call relations. LibD depends upon the directory structure of applications, which leads to the possibility of LibD’s failure due to obfuscation in the directory structure.
        
        \item \textbf{DepClean \cite{soto2021comprehensive}} detects the presence of bloated dependencies in Maven artifacts. Given an input maven project, DepClean analyzes the bloat library dependencies through API member calls to identify the actual used libraries by the project. The final output from the tool is new maven POM file containing only the used bloat libraries along with a dependency usage report. Authors conduct a qualitative assessment of DepClean on 9,639 Java projects hosted on Maven Central comprising a total of 723,444 dependency relationships. Some key observations of the study were that it is possible to reduce the number of dependencies of the Maven projects by 25\%. Further, it was observed that when pointed out, developers were willing to remove bloated dependencies.
        
        \item \textbf{JIngredient \cite{ishio2016software}} detects the TPLs present in a project's JAR file and proposes their origin TPLs. Basically, given an input TPL (in .jar format) $z$, JIngredient identifies the source TPLs containing source files similar to those present in $z$. To determine the similarity in source code, JIngredient uses class names (as class signatures) for classes present in $z$. As the tool depends on class names, which can easily be obfuscated, it is not resilient to source code obfuscations (also reported by the authors). JIngredient when compared with an existing software Bertillonage technique \cite{davies2013software}, it achieves an improvement of 64.2\% in terms of precision metrics. 
        
        \item \textbf{Autoflake \cite{autoflake}} removes unused TPL imports and unused variables from the code written in python. However, the removal of unused varables is disabled by default. Authoflake uses Pyflakes \cite{pyflakes} in its implementation. Pyflakes analyzes python programs to detect errors present in them  by parsing them.
        
        \item \textbf{PyCln \cite{pycln}} is another tool used to detect and remove unused TPL import statements in python source code.
     \end{enumerate}
 \end{enumerate}
%\pagebreak 
\textbf{Limitations of the current works:}
While the TPL-detection based on \enquote{reference list} methods tend to be inefficient and costly, the feature-based methods  are potentially affected by various types of obfuscations and are mostly developed for Android applications. Therefore, it is desirable to develop TPL-detection techniques that are resilient against such issues. Also, while surveying the literature, we observed that there are very few TPL detection tools for Python, and most of the solutions exist for Java or Android based applications. Therefore, there exist a need to develop efficient, obfuscation-resilient TPL detection tools for applications developed in other programming languages, such as Python, C\#, etc. To the best of our knowledge, we were not able to find any tools that detect TPLs for Python-based applications, i.e., applications having source code written in Python. AutoFlake and PyCln were the closest tools related to our work as they too work with TPLs, though they detect the bloatness or Irrelevance in terms of the usage of TPLs. In contrary, XtraLibD aims to detect the irrelvant TPLs by comparing with a certain collection of white-listed TPLs. Nevertheless, XtraLibD provides a novel solution for detection of irrelevant TPLs present in python applications.    % Thus, feature-based methods

\begin{table}[t]
\caption{Table of Notations.}
\centering 
    \begin{tabular}{r c p{11cm} }
    \toprule

    $T$ & $\triangleq$ & A TPL in its JAR, RAR, or ZIP file format.\\
    
    $L$ & $\triangleq$ & Set of considered programming languages, \{Java, Python\}.\\
    
    $C$ & $\triangleq$ & The collection of TPLs files fetched from MavenCentral or PyPi.\\
    
    $Z$ & $\triangleq$ & The set of PVA tuning-parameter variation scenarios, listed in Table-~\ref{table:training-scenarios}.\\

    $F_{bc}^{j}, F_{sc}^{j}, F_{bc}^{p}, F_{sc}^{p}$ & $\triangleq$ & The collections of bytecode ($bc$) and source code ($sc$) data obtained by disassembling and decompilation of .class and .pyc files $f$, respectively,  such that $f \in T$, and $T \in C$. Note: $j$ refers to Java (j) and $p$ refers to Python (p).\\

    $M_{bc}^{j}, M_{sc}^{j}, M_{bc}^{p}, M_{sc}^{p}$ & $\triangleq$ & The collections of Lib2Vec models trained on $F_{bc}^{j}, F_{sc}^{j}, F_{bc}^{p}, F_{sc}^{p}$,  $\forall Z$.\\
    
    $\hat{M_{bc}^{j}}, \hat{M_{sc}^{j}},\hat{M_{bc}^{p}}, \hat{M_{sc}^{p}}$ & $\triangleq$ & The best performing Lib2Vec models among all $M_{bc}^{j}, M_{sc}^{j}, M_{bc}^{p}, M_{sc}^{p}$.\\

    $\phi_{bc}^{j}, \phi_{sc}^{j}, \phi_{bc}^{p}, \phi_{sc}^{p}$ & $\triangleq$ & The PVA vectors corresponding to source files' bytecode and source code.\\
    
    $\hat{\phi_{sc}^{j}}, \hat{\phi_{bc}^{j}}, \hat{\phi_{sc}^{p}}, \hat{\phi_{bc}^{p}}$  & $\triangleq$ & The reference PVA vectors for source code and bytecode representations.\\

    $D$ & $\triangleq$ & The database containing the source files' vectors ($\phi_{bc}^{j}, \phi_{sc}^{j}, \phi_{bc}^{p}, \phi_{sc}^{p}$) for $C$.\\
   
   $\beta$ & $\triangleq$ & The number of training iterations or epochs used for training a PVA model.\\
    
    $\gamma$ & $\triangleq$ &  The PVA vector size used for training a PVA model. \\

    $\psi$ & $\triangleq$ & The number of training samples used for training a PVA model. \\

    $\alpha$ & $\triangleq$ & The cosine similarity score between two PVA vectors.\\
    $\hat{\alpha}$ & $\triangleq$ & The threshold cosine similarity score.\\
    
    \bottomrule
    \end{tabular}
\label{tab:BloatLibD-TableOfNotation}
\end{table}

\section{Proposed Approach}
\label{sec:BloatLibD-approach}
Our system's primary goal can be stated as follows: Given a TPL $T$, determine if $T$ is likely-to-be-irrelevant-TPL or a non-irrelevant-TPL in the given application. Our approach's central idea is to look for source code similarity and bytecode similarity between $T$ and the set of \enquote{available TPLs.} However, analyzing the detailed usages of the TPLs in the application is currently out of scope of this work. Our method can be considered as similar to the \enquote{reference list} methods, but the similarity here is determined on the basis of source code present in the TPL, and not merely the TPL names or package-hierarchial structure.  Table-~\ref{tab:BloatLibD-TableOfNotation} shows the notation used for various terms in this paper.

\subsection{Steps of our approach}
\label{sec:design-decisions}

The central ideas behind our approach were presented in Section~\ref{sec:basic-idea}. Here we expand those steps in more detail and highlight the relevant design decisions to be addressed while implementing each step.

\pagebreak

\begin{enumerate}
    \item \textbf{Preparing the dataset of \enquote{available TPLs}}
    \begin{enumerate}
        \item Download a set of Java TPLs $C$ from MavenCentral, and Python TPLs from Pypi\footnote{\url{https://pypi.org/}}.
        
        \emph{Design decision: Why use MavenCentral or Pypi to collect TPLs? How many TPLs should be collected from different software categories?}
        
        \item For each TPL $J \in C$, obtain the Java or Python source code and bytecode collections ($F_{sc}, F_{bc}$) by performing the decompilation and disassembly transformation operations.
        
        \emph{Design decision: Why are the decompilation and disassembly transformations appropriate?}
        
        \item Train the PVA models $M_{sc}$ and $M_{bc}$ on $F_{sc}$ and $F_{bc}$, respectively, obtained in the previous step.

        \emph{Design decision: Why use PVA, and what should be the PVA tuning-parameters for obtaining optimal results in our task?}
        
        \item For each source file $f \in F_{sc}$ and the bytecode record $b \in F_{bc}$, obtain the corresponding vector representations ($\phi_{sc},\phi_{bc}$) using suitable PVA models trained in the previous step. $\phi_{sc}$ and $\phi_{bc}$ obtained for each source code and bytecode instance are stored in the database $D$. 
       
    \end{enumerate}
    
    \item \textbf{Determining if an input TPL ($T$) is a irrelevant-TPL or not for a given application}
    \begin{enumerate}
        \item Compute the vector representation $\langle \phi_{bc}', \phi_{sc}' \rangle$  for the bytecode and source code representations of $T$.
        \item Obtain all the vectors  $\langle \phi_{bc}, \phi_{sc} \rangle \in D$, such that the respective similarity scores between $\langle \phi_{bc}', \phi_{bc} \rangle$ and $\langle \phi_{sc}', \phi_{sc} \rangle$ are above specific threshold values ($\hat{\alpha_{bc}}$ and $\hat{\alpha_{sc}} $).
        
        \emph{Design decision: What are the optimal  values of similarity thresholds ($ \hat{\alpha_{bc}}$ and $\hat{\alpha_{sc}} $)?}
        
        \item Determine whether $T$ is a irrelevant-TPL or not for the given application.
        
        \emph{Design decision: How is the nature of $T$ determined?}
    \end{enumerate}

\end{enumerate}

%\pagebreak
\subsection{Design considerations in our approach}
\label{sec:design-decisions}

In this section, we address the design decisions taken while implementing our approach. 

\subsubsection{Collecting TPLs from MavenCentral and Pypi:}
\label{sec:algo-step-collect-libs}
The libraries used for training our models (named Lib2Vec) were taken from MavenCentral and Pypi software repository hosts. We use MavenCentral to fetch Java-TPLs and Pypi for Python-TPLs.  We choose these portals as these are the public hosts for a wide variety of popular Java and Python libraries. Further, MavenCentral categorizes the Java libraries based on the functionality provided by the libraries. However, our method is not dependent on MavenCentral or Pypi; the TPLs could be sourced from reliable providers.  To collect the TPLs, we perform the following steps:
    \begin{enumerate}
        \item \label{step:download-jar} Crawl the page \texttt{\url{https://mvnrepository.com/open-source}} and \texttt{\url{https://pypi.org/}}, and download the latest versions of TPLs in JAR and .rar (or .zip) formats, respectively. We download \underline{top k (=3) libraries} listed under \underline{each category} at MavenCentral. Similarly, we downloaded a random collection of python TPLs from PyPi. Pypi categorizes the TPLs based on different programming languages and frameworks used while developing them. While collecting the TPLs, we made sure to download Python libraries belonging to different frameworks to obtain a heterogeneous TPL dataset. Further, we applied the following constraints to obtain a useful (non-trivial) collection of TPLs:
        \begin{enumerate}
            
        \item \emph{Size constraint:} The size of library should be greater than a threshold ($\geq9$ KB). Please note that repository size here stands for the total size of only the source files present in the repository. All the binary files such as multimedia and libraries are excluded.

        \item \emph{Source file count constraint:} The repository should contain at least one source file written in Java or Python.
        
        \item \emph{Reputation constraint:} The repository should have earned $>=1$ star. This constraint was applied to ensure that the selected repositories are popular and are being used by a minimum number of programmers.
         \end{enumerate}
        \item \label{step:extract-metadata} Extract and save in a database table the metadata associated with the downloaded TPLs. The metadata includes details of the TPL, such as the category, tags, and usage stats.
    \end{enumerate}

\subsubsection{Rationale for choosing PVA for training models:}
\label{sec:building-models}

We train Lib2Vec models using PVA on the source code and bytecode textual forms of the .class files obtained by the decompilation and disassembly of various JAR TPLs. In case of Python-TPLs (available in .zip or .rar formats), we directly obtain the source files by uncompressing them. We then obtain the bytecode by first performing the compilation and then the disassembly process as shown in Fig.~\ref{fig:basic-idea}, and discussed in Section~\ref{sec:basic-idea}. For our experiments, we train language-specific and type-specific PVA models, i.e., a Lib2Vec model trained on python source code $M_{sc}^{p}$, a Lib2Vec model trained on python bytecode $M_{bc}^{p}$, and similarly for Java ($M_{sc}^{j}, M_{bc}^{j}$).  The key reasons for choosing PVA are i) It allows us to compute the fixed-length vectors that accurately represent the source code samples. Keeping the length of vectors same for every source code sample is critical for implementing an efficient and fast system. ii) Recent works \cite{Code2VecAlon2019} propose a close variant of PVA, and have proven that it is possible to compute accurate vector representations of source code and that such vectors can be very useful in computing semantic similarity between two source code samples.  

\textbf{Tuning parameters for PVA:} Performance, in terms of accuracy, efficiency, and speed of PVA, is determined by its input parameters such as $\beta, \gamma$, and $\psi$ (see Table-~\ref{tab:BloatLibD-TableOfNotation}). Therefore, one of the major tasks is to select the optimal values of $\beta, \gamma$, and $\psi$ that can result in the best performing Lib2Vec models ($\hat{M_{bc}^{j}}, \hat{M_{sc}^{j}}, \hat{M_{bc}^{p}}, \hat{M_{sc}^{p}}$). The experiments' details to determine $\beta, \gamma$, and $\psi$  are provided in the Appendix. 

\subsubsection{Rationale for using the decompilation and disassembly transformations:}
\label{sec:decomp-disassembly-process}
It is necessary to derive a \enquote{normalized} and obfuscation-resilient textual form of the source files to compute a reliable vector representation. The normalization applies a consistent naming of symbols while preserving the semantics and structure of the code. We use the decompilation (\textit{giving a source code text}) and disassembly (\textit{giving a bytecode text}) as transformations to extract such normalized textual forms of .class (or source) files.

\subsubsection{Employing the use of vector representations for performing similarity detection between TPLs:}
\label{sec:algo-step-whitelist-db}
To determine the similarity  between libraries efficiently, we create a database (D) of vectors.  These vectors correspond to the source files present in a target repository of libraries (such as MavenCentral, or an in-house repository maintained by an organization). We obtain the vector representations for both the source code and bytecode of source files present in TPLs using suitably trained PVA models and store them in D. The PVA vectors enable fast and efficient detection of TPL similarity. % Algorithm~\ref{alg:storing-pv-vectors} lists the significant steps involved. 

\subsubsection{Computing the threshold similarity measure $\hat{\alpha}$}
\label{sec:threshold-similarity}
Our method detects two libraries' similarity by inferring the similarity scores for source files contained in those libraries. To check if two source file vectors  are similar or not, we compute their cosine similarity\footnote{\url{https://bit.ly/2RZ3W5L}}. 
An important design decision in this context is:

\emph{For reliably detecting a library, what is the acceptable value of Lib2Vec similarity scores for decompiled source code and bytecode inputs?}

We deem two source files (.java or .py files) as highly similar or identical when the similarity score for the files is higher than a threshold value $\hat{\alpha}$. Note: In these comparisons, we compare the files written in the same language only. The value of $\hat{\alpha}$ is determined by running several experiments to measure similarity scores for independent testing samples. The details of the experiments are discussed in the Appendix.

\subsubsection{Determining the nature of an unseen TPL $T$ for a given application $A$:}
\label{sec:algo-step-detect}
To determine if a given TPL ($T$) is \enquote{irrelevant-TPL} for an application ($A$), we leverage the  best performing Lib2Vec models ($\hat{M_{bc}}$ and $\hat{M_{sc}}$) and the vectors database $D$. 

If $L$ contains .class or .py files depicting considerable similarity ($\geq\hat{\alpha}$) with the \enquote{available TPLs,} it is deemed to be as \enquote{likely-to-be-relevant} for $A$. If for at least $N$ source files in $T$, the similarity scores are $\geq\hat{\alpha}$, we label $T$ as a \emph{likely-to-be-relevant} TPL for $A$, else a \emph{irrelevant-TPL}. In our experiments, we take N as half of the count of source files present in $T$. For a more strict matching, a higher value of N can be set. %The $\hat{\alpha}$ values listed for bytecode and source code are listed in Table~\ref{table:thresholds}. 
The complete steps for the detection procedure are listed in Algorithm~\ref{alg:testing-unseen-JAR}. 

\emph{Selection of the top-similar \enquote{available TPLs}:} 
We explain it with an example. Suppose we have four \enquote{available Java TPLs}, with $C:=$ \{M1.jar, M2.jar, M3.jar, M4.jar\}, such that these contain 15, 6, 10, and 10 .class files, respectively. Now, $D$ will contain the PVA vectors corresponding to the source code and bytecode representations of all the .class files present in all the JARs in $C$. Next, suppose we want to test a JAR file Foo.jar that contains ten .class files, and that we have the following similarity scenarios: 
\begin{enumerate}
    \item All ten .class files of \emph{Foo.jar} are present in M1.jar.
    \item All six .class files of \emph{M2.jar} are present in Foo.jar.
    \item Seven out of ten .class files of \emph{M3.jar} are present in Foo.jar.
    \item For \emph{M4.jar}, none of these files is identical to those present in Foo.jar, but they have similarity scores higher than the threshold. 
\end{enumerate}

\emph{Which of the JAR files (M1 -- M4) listed above will be determined as the most similar to Foo.jar?} 

Our approach determines the most-similar JAR file by measuring the total number of distinct .class file matches. So with this logic, the similarity ordering for Foo.jar is M1, M4, M3, M2.

In this setting, determining the similarity of two JARs is akin to comparing two sets for similarity. Here the items of the \emph{sets} would be the PVA vectors representing .class files. We apply the following approach to determine the TPL-similarity: 
\begin{enumerate}
    \item For each .class $c$ in Foo.jar, find each record $r \in D$ such that the similarity score between $c$ and $r$ is higher than a threshold. Let $R\subset D$ denote the set of all such matched records.
    \item Find the set $Y$ of distinct JARs to which each $r\in R$ belongs.
    \item Sort $Y$ by the number of classes present in $R$.
    \item Select the top-k items from $Y$ as similar JARs to Foo.jar.
\end{enumerate}
Algorithm-~\ref{alg:testing-unseen-JAR} presents the above logic in more detail. The same algorithm works for both the JAR-based and Python-based TPLs.

\begin{algorithm}
  \small
  \caption{\small{Steps for determining the nature of a TPL $T$.}}
  \label{alg:testing-unseen-JAR}
 \small{
  \begin{algorithmic}[1]
    \STATE \textbf{Input:} $T := $ A TPL file provided as input by an end-user.\\
    $L := $  Set of considered programming languages, \{Java, Python\}\\
    $\hat{M_{bc}^{j}}, \hat{M_{sc}^{j}},\hat{M_{bc}^{p}}, \hat{M_{sc}^{p}} := $ The best performing Lib2Vec models. \\ 
    
    $\hat{\alpha_{sc}^{j}}, \hat{\alpha_{bc}^{j}},\hat{\alpha_{sc}^{p}}, \hat{\alpha_{bc}^{p}} := $ Threshold similarity scores for source code and bytecode for Python and Java.\\
    $\hat{\phi_{sc}^{j}}, \hat{\phi_{bc}^{j}}, \hat{\phi_{sc}^{p}}, \hat{\phi_{bc}^{p}} := $ Reference PVA vectors for source code and bytecode for Python and Java.\\
    $D :=$ Database containing the vector representations of source files in C.\\
    \STATE \textbf{Output:} $d :=$ Decision on the nature of $J$.\\
    
    /*Please see Table~\ref{tab:BloatLibD-TableOfNotation} for notation*/\\
    \STATE $S_{sc} := S_{bc} := NULL$
    
   % \textbf{Steps:}
    \FORALL{.class files or .py files $f \in T$}
       \STATE l := Detect Programming Language of f. 
        \STATE Obtain the PVA vectors $\langle \phi_{bc}^{l}, \phi_{sc}^{l} \rangle$ for $\langle f_{bc}, f_{sc} \rangle$ using  $\langle \hat{M_{bc}^{l}}, \hat{M_{sc}^{l}} \rangle$, where $l\in L$. %/* As per Algorithm~\ref{alg:storing-pv-vectors}*/
            \STATE \label{step:compute-cos-sim} Query the database $D$ for top-k most similar vectors to $\phi_{sc}^{l}$ and $\phi_{bc}^{l}$. 
            \STATE $\alpha_{sc}^{l}, \alpha_{bc}^{l} := $ Compute the cosine similarity between $\langle \phi_{sc}^{l}, \hat{\phi_{sc}^{l}} \rangle $ and  $\langle \phi_{bc}^{l}, \hat{\phi_{bc}^{l}} \rangle$.
            \STATE $S_{sc} := S_{sc} \cup \langle \alpha_{sc} \rangle$
            \STATE $S_{bc} := S_{bc} \cup \langle \alpha_{bc} \rangle$  
   \ENDFOR
  \STATE \[  d :=\begin{cases}
       \text{relevant} &\parbox[t]{0.35\textwidth}{if for at least $N$ source file records 
       in both $S_{sc}$ and $S_{bc}$ individually,
       $\alpha_{sc}^{l} > \hat{\alpha_{sc}^{l}}$ and   $\alpha_{bc}^{l} > \hat{\alpha_{bc}^{l}}$ respectively.}\\
       \text{irrelevant} &\quad\text{otherwise}\\
     \end{cases}\]

  \end{algorithmic}}
\end{algorithm}

 \begin{figure}
\centering
    \includegraphics[width=0.8\columnwidth]{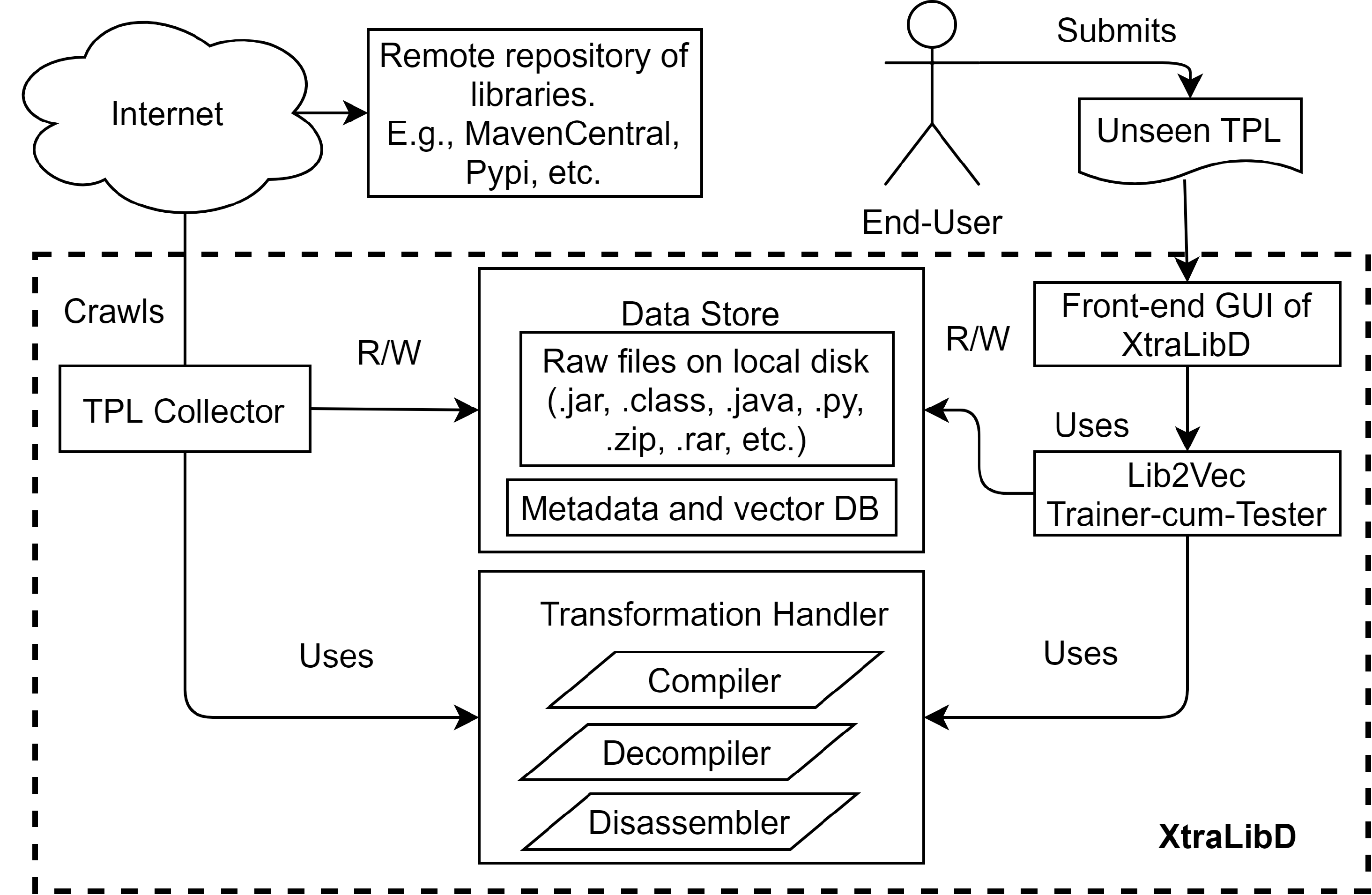}
    \caption{Logical structure of the proposed system \cite{dewan2021bloatlibd}.}
    \label{fig:logical_structure}
\end{figure}

\subsection{Implementation details}
\label{sec:BloatLibD-impl}

The logical structure of the proposed system is shown in Figure~\ref{fig:logical_structure}. All components of the system have been developed using the Python programming language. Details of each of the components are as follows:

\begin{enumerate}
    \item \textbf{TPL file collector:} \label{component:jar-collector} We developed a crawler program to collect the TPLs and the metadata associated with each file. The files were downloaded from \texttt{\url{www.mavencentral.com}} and \texttt{\url{https://pypi.org/}}, reputed public repositories of Open Source Software (OSS) Java and Python libraries. MavenCentral has about 15 million indexed artifacts, which are classified into about 150 categories. Some examples of the categories include JSON Libraries, Logging Frameworks, and Cache Implementations. 
    The metadata about each JAR includes the following main items: License, Categories, HomePage, Date of Availability, Files, Used By (count, and links to projects using a library). %The JAR file categories' complete details, metadata collected, and the specific JAR files chosen can be found at \textit{\url{https://bit.ly/2WFALXf}}.
    
    Similarly, Pypi has $300+K$ python projects classified by programming languages, topics, frameworks, etc., used while developing them. Django, Dash, CherryPy, Flask, and IDLE are some of the example Pypi frameworks.  
    The metadata about each Python project on Pypi comprises release-history, project readme, the count of stars, forks, pull requests, Date of Release, Latest Version, HomePage, License, Author information, Maintainers information, Programming Environment, Programming Framework, Intended Audience, Operating System, and Programming Language used.

    \item \textbf{Transformation handler:} \label{component:transformation-handler} This module provides the transformations and preprocessing of the source files present in the input TPL files (in JAR or .zip/.rar formats). Three types of transformations implemented are a) Decompilation of the .class file present in JAR files to produce a corresponding Java source, b) Compilation of the .py files present in .rar/.zip Python TPLs to produce the corresponding compiled .pyc files, and c) Disassembling the .class and .pyc files into human-readable text files containing bytecode instructions for the .class files and .pyc files, respectively. 
    
    We used the Procyon \cite{procyon} tool for performing the decompilation and disassembling of the .class files. The respective transformation output is further preprocessed to remove comments and adjust token whitespaces before storing it as a text file in a local repository. The preprocessing was done for the decompiled Java source to ensure that the keywords and special tokens such as parentheses and operators were delimited by whitespace. Similar, we removed the comments from the python source code (in .py) files during the preprocessing phase. The preprocessing provides proper tokenization of the source into meaningful \enquote{vocabulary words} expected by the PVA.
    
    \item \textbf{Lib2Vec trainer-cum-tester:} \label{component:Lib2Vec-trainer-cum-tester} We use an open-source implementation of the PVA -- called Gensim \cite{par-vec-nips15}, to train our  Lib2Vec models. The Lib2Vec trainer-cum-tester module's primary task is to: 
\begin{enumerate}
    \item Train the Lib2Vec models using bytecode and source code produced by various disassembling, compilation, and decompilation operations.
    
    \item Infer the vectors for unseen .class files' bytecode and source code by using the respective models.

\end{enumerate}
%It implements the methodology steps outlined in Algorithms~\ref{alg:testing-unseen-JAR},~\ref{alg:training-pv-model},~\ref{alg:procedure-to-test-models}, and ~\ref{alg:storing-pv-vectors}.

\item \textbf{Metadata and the vectors' database:} \label{component:Lib2Vec-metadata-vec-db} The information about libraries fetched from MavenCentral is stored in a relational database. The following are the essential data items stored in the database:
\begin{enumerate}
    \item Name, category, version, size, and usage count of the library.
    \item Location of the library on the local disk as well as a remote host.
    \item For each .class file $f$ in a JAR or .py file $f$ in a Python project:
    \begin{enumerate}
        \item The fully qualified name of the $f$.
        \item Sizes of $f$, and the textual form of its decompiled Java source code ($f_{sc}^{j}$), python source code ($f_{sc}^{p}$), and the disassembled bytecode for both Java and Python TPLs ($f_{bc}^{j},f_{bc}^{p}$).
        \item Inferred PVA vectors $\langle \phi_{sc}^{l}, \phi_{bc}^{l}\rangle$, $l\in L$ for the above files.
         \item Cosine similarity scores $\hat{\alpha_{sc}^{l}}$ and $\hat{\alpha_{bc}^{l}}$ between $\langle \phi_{sc}^{l}, \hat{\alpha_{sc}^{l}} \rangle$ and $\langle \phi_{bc}^{l}, \hat{\alpha_{bc}^{l}}\rangle$, respectively.
        The values $\hat{\alpha_{sc}^{l}}$ and $\hat{\alpha_{bc}^{l}}$ are scalar.
    \end{enumerate}
\end{enumerate}

\begin{figure}
\centering
    \includegraphics[width=0.8\columnwidth]{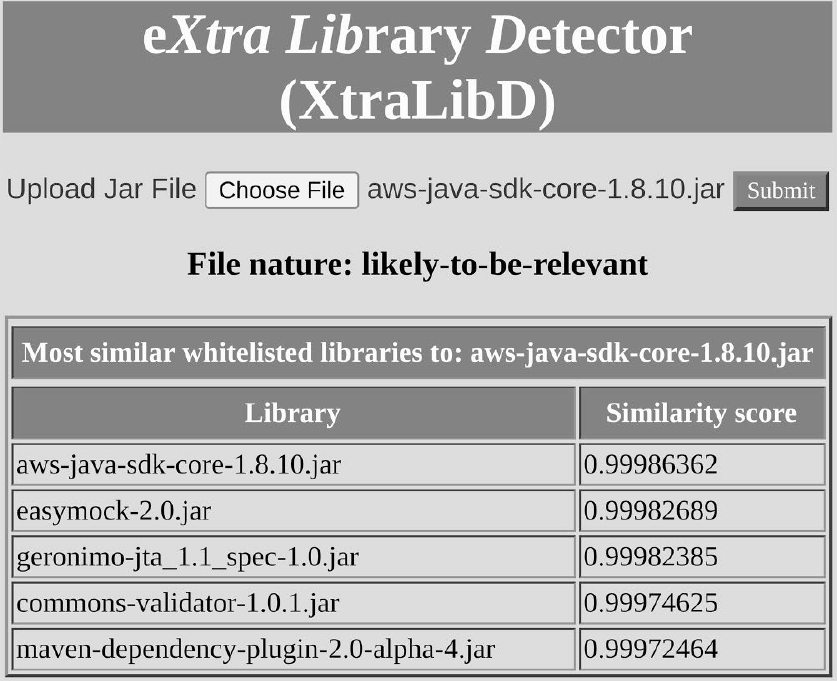}
    \caption{Top similar TPLs detected by XtraLibD.}
    \label{fig:front-end-BloatLibD}
\end{figure}

\item \textbf{XtraLibD's GUI:} \label{component:BloatLibD-GUI}
The user interface of XtraLibD is a web-based application.  End-user uploads a TPL using this GUI, which is then processed by our tool at the server-side. The tool requires the TPLs in JAR, .zip, or .rar formats as input. Figure~\ref{fig:front-end-BloatLibD} displays the results for a test file\footnote{\url{https://bit.ly/2yb2eHY}} submitted to our tool. As shown by the figure, XtraLibD displays the input file's nature and the top-k (k=5) similar essential libraries along with the corresponding similarity scores. As we achieve higher accuracy with the source code Lib2Vec models than the bytecode models (discussed in the Appendix), we use the best performing source code Lib2Vec model for developing our tool. 
 
\end{enumerate}

\section{Experimental Evaluation}
\label{sec:BloatLibD-experiments}

The primary goal of our experiments is to validate the correctness and accuracy of our tool -- XtraLibD. For a given input file, XtraLibD labels it as $\langle$ likely-to-be-irrelevant-TPL, likely-to-be-non-irrelevant-TPL$\rangle$, and lists the top-k similar libraries and the respective similarity scores shown in Figure~\ref{fig:front-end-BloatLibD}.  The efficacy of our tool depends on its accuracy in performing the task of detecting similar TPLs. XtraLibD achieves this by detecting the similarity between the PVA vectors of the .class files present in the TPLs. The Lib2Vec models used by XtraLibD are responsible for generating different PVA vectors. %Thus, the accuracy of the TPL-detection task depends on the accuracy of the Lib2Vec models. 
Therefore, we perform various parameter-tuning experiments to obtain the best performing Lib2Vec models (discussed in the Appendix). To evaluate the performance of XtraLibD, we develop a test-bed using the TPLs collected from MavenCentral and PyPi (discussed in Section~\ref{sec:test-bed-setup}) and perform the following experiments:
\begin{enumerate}
\item Test the performance of Lib2Vec models (and thus XtraLibD) in performing the TPL-detection task (discussed in the Appendix). 
\item Compare the performance of XtraLibD with the existing TPL-detection tools (discussed in Section~\ref{sec:comparison-with-existing-tools}).
\end{enumerate}

\subsection{Test-bed setup}
\label{sec:test-bed-setup}
To compare the performance of XtraLibD on Java-based and Python-based tools, we setup Java and Python testbeds separately for our experiments. In this section, we provide the deatils of setting these testbeds.

\subsubsection{Developing Test-bed for comparison with Java-based tools:}
We crawled \texttt{\url{https://mvnrepository.com/open-source?p=}\textbf{PgNo}}, where \textbf{PgNo} varied from 1 to 15. Each page listed ten different categories from the list of most popular ones, and under each category, the top-three libraries were listed.

We started by downloading one JAR file for each of the above libraries. That is, a total of $15 \times 10 \times 3 = 450$ JAR files were fetched. In addition to the above JAR files, we also included the JDK1.8 runtime classes (\texttt{rt.jar}). After removing the invalid files, we were left with 97 JAR files containing 38839 .class files. 

We chose random 76 JAR files out of 97 plus the \texttt{rt.jar} for training the Lib2Vec models, and the remaining 20 JAR files were used for testing. We used only those .class files for training whose size was at least 1kB since such tiny .class files do not give sufficient Java and byte code, which is necessary to compute a sufficiently unique vector representation of the .class contents.
The training JARs had 33,292 .class files, out of which only 30,427 were of size 1kB or bigger. The testing JARs had 4,033 .class files. 

\subsubsection{Developing Test-bed for comparison with Python-based tools:} We crawled \texttt{\url{https://pypi.org/search/?c=Programming+Language+\%3A\%3A+Python&o=&q=&page=PgNo}}, where \textbf{PgNo} varied from 1 to 500. Since Pypi also categorizes the projects based on the \textit{Developing Framework}, we made sure to select top 10 libraries belonging to each framework on different pages, resulting in a total of 43,711 python TPLs. By applying the constraints during the selection of these libraries (discussed in Section~\ref{sec:design-decisions}), we made sure that each library contains atleast one source file written in python. These libraries were available in .zip or .rar formats, and by uncompressing them we obtained 42,497,414 .py files.
Out of the total 42,497,414 .py files, only 682,901 .py had size $>=1$kB, which we considered for training and testing our python Lib2Vec models. From these 682,901 .py files, we choose random 30,598 .py for training and rest 13,113 for test our python Lib2Vec models.  

Note: We chose the minimum source file size as 1kB because we observed that the files smaller than 1kB did not significantly train an accurate Lib2Vec model. A summary of the TPL data  is shown in Table~\ref{table:dataset}. Note: the training and testing of Lib2Vec models were performed on the source code and bytecode extracted from the respective number of source files.

\begin{table}[t]
\centering
\caption{TPL Data summary.}
\label{table:dataset}
 \resizebox{0.9\columnwidth}{!}
    {
\begin{tabular}{ |p{5cm}| l| l|}
\toprule
\textbf{Item} & \textbf{Java Count \cite{dewan2021bloatlibd}} & \textbf{Python Count} \\ 
\midrule
Downloaded TPLs & 450 & 43,711 \\ \hline
TPLs selected for experiments & 97 & 43,711 \\ \hline 
TPLs used for training & 76 + 1 (\texttt{rt.jar}) & 30,597+ 1 (\texttt{rt.zip}) \\ \hline 
TPLs used for testing & 20 & 13,113\\ \hline
.class (or .py) files used for training & 30427 & 478,031\\ \hline 
.class (or .py) files used for generating test pairs of bytecode and source code & 4033 & 204,870\\ \hline
Unique pairs of bytecode files used for testing & 20,100 & 200,000 \\ \hline 
Unique pairs of source code files used for testing & 20,100 & 200,000\\ 
\bottomrule
\end{tabular}}
\end{table}

\subsection{Performance Comparison of XtraLibD with the existing TPL-detection tools}
\label{sec:comparison-with-existing-tools}
To the best of our knowledge, no work leverages the direction of using code similarity (in TPLs) and the vector representations of code to detect the irrelevant-TPLs for Java and Python applications. We present our tool's performance comparison (XtraLibD) with some of the prominently used tools, viz., LiteRadar, LibD, and LibScout, DepClean, JIngredient, Autoflake, and PyCln. The details about these tools have been discussed in Section~\ref{sec:literature-review-TPL-tools}. We already conducted the experiments with LiteRadar, LibD, and LibScout in our previous research work \cite{dewan2021bloatlibd}, and extend our previous work by providing the TPL-detection support for python-based TPLs. In this section, we provide the details of new experiments perfomed with some of the recent Java-based and Python-based TPL detection tools, viz., DepClean, JIngredient, Autoflake, PyCln, and also a summarized comparison of performance results obtained in our previous comparisons \cite{dewan2021bloatlibd} with LiteRadar, LibD, and LibScout.

Though PyCln and Autoflake work in detection of used/ unused import TPLs, and not the detection of TPLs, we included them as in our research, we found them to be the closest available python tools working with TPLs. For experiment with PyCln and Autoflake we experimented by developing different import scenarios, for instance, direct imports, secondary imports, and both with used and unused cases.

\subsubsection{Objective} 
To compare the performance of XtraLibD with the existing TPL-detection tools. Through this experiment, we address the following:

\emph{How does XtraLibD perform in comparison to the existing TPL-detection tools? What is the effect on storage and response time? Is XtraLibD resilient to the source code obfuscations?}

\subsubsection{Procedure}
To perform this experiment, we invited professional programmers and asked them to evaluate our tool. One hundred and one of them participated in the experiment. We  had  a  mixture  of programmers  from  final  year  computer  science  undergraduates, postgraduates,  and  the  IT  industry  with  experience  between 0-6  years. The participants had considerable knowledge of Java programming language, software engineering fundamentals, and several Java and Python applications. The experiment was performed in  a controlled industrial environment. We provided access to our tool for performing this experiment by sharing it at \url{https://www.doi.org/10.5281/zenodo.5179747}. The tools' performance was evaluated based on their \emph{accuracy}, \emph{response time}, and the \emph{storage requirement} in performing the TPL-detection task.  We compute the tool's storage requirement of the tools by measuring the memory space occupied in storing the relevant \enquote{reference TPLs.} The TPL-detection tools -- LibD, LibScout, and LibRadar, require the inputs in an Android application PacKage (APK) format. Therefore, APK files corresponding to the JAR versions of the TPLs were generated using the Android Studio toolkit\footnote{https://developer.android.com/studio} (listed in Step~\ref{step:conversion-to-APK} of Algorithm~\ref{alg:comp-tools}). Similarly, DepClean requires the input TPLs in maven project format and JIngredient in JAR file format. Both PyCln and Autoflake require the input files in .py format.

The programmers were requested to perform the following steps:

\begin{enumerate}
    \item Randomly select a sample of 3-5 TPLs from the test-bed developed for the experiments (discussed in Section\ref{sec:test-bed-setup}).
    \item Test the TPLs using Algorithm~\ref{alg:comp-tools}.
    \item Report the tools' accuracy and response time, as observed from the experiment(s). 
\end{enumerate}

\begin{algorithm}
  \caption{Steps for performing the comparison.}
  \label{alg:comp-tools}
  \small{
  \begin{algorithmic}[1]
    \STATE \textbf{Input:} $C = $ Set of TPLs ($T$) downloaded  randomly from MavenCentral and Pypi.\\
    $X = $ Set of XML files required as input by LibScout. \\
    $\hat{M_{bc}^{j}}, \hat{M_{sc}^{j}, M_{bc}^{p}}, \hat{M_{sc}^{p}} := $ The best performing Lib2Vec models for Java and Python. \\ 
   $\hat{\phi_{sc}^{j}}, \hat{\phi_{bc}^{j}}, \hat{\phi_{sc}^{p}}, \hat{\phi_{bc}^{p}} := $ Reference PVA vectors for source code and bytecode.\\
    /*Please see Table~\ref{tab:BloatLibD-TableOfNotation} for notation*/\\
    \STATE \textbf{Output:} 
    Terminal outputs generated by the tools.\\
%    \textbf{Steps:}
    \STATE $F_{sc}^{p} := F_{bc}^{p} := F_{sc}^{j} := F_{bc}^{j} := NULL$
    \FORALL{TPLs $T \in C$}
        \FORALL{.class and .py files $f \in T$}
           \STATE l := Detect the programming language of $f_u$.
            \STATE $f_{bc}^{l}, f_{sc}^{l}  := $ Obtain the textual forms of the bytecode and source code of $f_l$.
            \STATE $f_{bc}', f_{sc}'  := $ Modify $f_{bc}^{l}$ and $ f_{sc}^{l}$ using transformations listed in Section~\ref{sec:test-bed-setup}.
            \STATE $F_{sc}^{l} := F_{sc}^{l} \cup \langle f_{sc}' \rangle$
            \STATE $F_{bc}^{l} := F_{bc}^{l} \cup \langle f_{bc}' \rangle$
        \ENDFOR
    \ENDFOR
    \STATE $Y := $\label{step:conversion-to-APK} Convert $F_{sc}^{l}$ into the respective input formats required by the tools. 
    \STATE Test with the considered tools using $X$ and $Y$.
    \STATE Test $F_{sc}^{l}$ and $F_{bc}^{l}$ with XtraLibD using Algorithm~\ref{alg:testing-unseen-JAR}.
  \end{algorithmic}}
\end{algorithm}

\subsubsection{Evaluation criteria} In the context of the TPL-detection, we define the accuracy as \cite{dewan2021bloatlibd}:

\noindent
\begin{align}
    \label{eq:accuracy-of-BloatLibD}
    Accuracy= \frac{Number\ of\ TPLs\ correctly\  detected}{Total\ number\ of\ TPLs\ tested}
\end{align}

Similarly, for the detection of used/ unused TPL imports, we define accuracy as follows:

\noindent
\begin{align}
    \label{eq:accuracy-of-BloatLibD}
    Accuracy= \frac{Number\ of\  correctly\  identified(used/ unused)\ TPL\ imports}{Total\ number\ of\ TPL\ imports}
\end{align}

By a correctly identified TPL import we imply to a scenario when a used import is labelled (or identified or marked) as a used import, and an unused import is labelled as an uned import by a considered tool.

\subsubsection{Results and observations}
Table~\ref{tab:response-time} lists the \emph{accuracy}, \emph{response time}, and \emph{storage space requirement} values observed for the tools. We now present a brief discussion of our results.

\textbf{Accuracy of the TPL-detection tools:}
Some of the key observations from the experiments are:
\begin{enumerate}
    \item LiteRadar cannot detect the transformed versions of the TPLs and fails in some cases when tested with the TPLs containing \emph{no transformations}. For instance, it cannot  detect exact matches in the case of  zookeeper-3.3.0.jar library\footnote{\url{http://bit.ly/2VymUmA}} and kotlin-reflect-1.3.61.jar library\footnote{\url{http://bit.ly/32MvkZe}}. 
    \item LibScout detects the \emph{TPLs without any transformations} but suffers from package-name obfuscations as it cannot detect the modified versions of TPLs containing package-name transformations. 
    \item LibD substantially outperforms LibRadar and LibScout in capturing the similarity between the TPLs but does not comment on their nature, i.e.,  $\langle$likely-to-be-irrelevant-TPL, likely-to-be-non-irrelevant-TPL$\rangle$. It also comes with an additional cost of manually comparing the TPLs with the \enquote{reference set.} 
    \item DepClean is able to detect 99\% of the direct dependency cases, but is unable to detect transitive dependency cases as observed by us and an existing research \cite{serena2021}.
    
    \item JIngredient detects the TPLs and their class reuse with an accuracy of 99.75\%. However, its dependence on class names to form the class signatures used to detect TPL similarity, makes it prone to obfuscations. Also, its use is constrained by huge storage and processing requirements (discussed shortly in detail). 
    
    \item  Since PyCln and Autoflake do not perform TPL detection and both perform the detection of used/ unused TPL imports, we analyze them independenly and compare them with each other. In our experiments, PyCln was unable to detect the import statements present in a try-except structure, whereas Autoflake was successful in detecting them. Further, both the tools miss the import statements where one library imports another but does not use all the models of it. For instance, when a file A.py imports from  B.py (as from B import *), and does not use all the modules imported in B.

    \item XtraLibD detects the TPLs for 99.5\% of the test cases. As observed from the table values, XtraLibD outperforms LiteRadar, LibScout, and LibD with the improvement scores of 30.33\%, 74.5\%, and 14.1\%, respectively. Though XtraLibD has a comparable accuracy rate as DepClean, JIngredient, Autoflake, and PyCln, its benefits in terms of low storage requirements and small response time make it more feasible for use.  Also, XtraLibD performs equally well on the obfuscated test-inputs, the results validate that it is resilient to the considered obfuscation types. 
    %the input TPLs and provides a similarity score $>$ 95\% for all the test cases.
\end{enumerate}

\begin{table}[t]
\centering
\scriptsize
\caption{Performance comparison of various TPL-detection tools.}
\label{tab:response-time}
\resizebox{0.9\columnwidth}{!}
    {
\begin{tabular}{|c| c| c| c|}
\toprule[1pt]
\multirow{3}{*}{\shortstack{TPL\\ detection\\ tools}}& \multicolumn{3}{c|}{Performance Metrics values}\\
\cline{2-4}
&\multirow{3}{*}{\shortstack{Accuracy\\ (in \%)}}&\multirow{3}{*}{\shortstack{Response Time\\ (in seconds)}} & \multirow{3}{*}{\shortstack{Storage requirement\\ (in MBs)}}\\
&&&\\
&&&\\
\midrule[1pt]

LiteRadar & 68.97 & 12.29 & 1.64 \\ \hline
LibScout & 25.23 & 6.46 & 3.93 \\ \hline
LibD & 85.06 & 100.92 & 12.59 \\ \hline
DepClean & 99 & 4.458 & Not applicable\\  \hline
JIngredient &  99.75 & 0.001 & 1000 \\  \hline
Autoflake & 99.29 & 0.002 & Not applicable \\  \hline
PyCln & 99.29 & 0.002 & Not applicable \\  \hline
XtraLibD (Java) & 99.48 & 12.043 & 1.52 \\ \hline
XtraLibD (Python) & 99.5 & 12.043 & 401.94 \\

\bottomrule[1pt]
\end{tabular}
}
\end{table}

%\pagebreak
\textbf{Storage requirement of the TPL-detection tools:} XtraLibD leverages the PVA vectors to detect the similarity among the TPLs, while the tools used for comparison, viz., LibD, LibScout, and LiteRadar, use the \enquote{reference lists} of TPLs. These tools contain the \enquote{reference lists} of TPLs as files within their tool packages. As observed from the storage requirement values, XtraLibD has the lowest storage requirement due to the vector representation format. XtraLibD reduces the storage requirement by 87.93\% compared to LibD, 61.28\% compared to LibScout, and 7.3\% compared to LiteRadar. As DepClean is based on the TPL dependency usage rather than the comparison with a reference list of libraries, storage requirement comparison is not applicable to it. Similarly, PyCln and AutoFlake are based on code analysis for TPL import detection, hence the storage requirement comparison is not applicable in their case. 

JIngredients performs TPL detection by using the source code class names as signatures to match with a reference database for determining TPL reuse. However, JIngredients has a storage requirement of 1 GB for its database, and needs a high performance hardware and memory support to implement its approach. It was originally implemented on a high end workstation with 2 Six core processors with a 64 GB RAM on a large corpus size of 1 GB ( which itself was  constructed from an original repository data of size 77.8 GB with a total of 172,232 JAR files). To perform our experiments, we used a database of 214 JAR files from which 192 JAR files were used in JIngredient's database. We performed the experiments on a machine having an AMD Ryzen 5 4600H 3.0 Ghz 6 Cores 12 Threads processor with an 8 GB RAM. JIngredients was unable to detect any instances of reuse on this machine, though it is able to identify the classes within the JAR files. Thus, JIngredient's working is constrained by its high storage and processing requirement. However, XtraLibD has only 1.52 MBs storage requirement for its Java version at a comparable accuracy rate as JIngredient. Thus, when compared in terms of storage requirement, XtraLibD outperforms JIngredient by 99.85\%. 

\textbf{Response time of the TPL-detection tools:} For DepClean, the average response time mentioned in Table~\ref{tab:response-time} includes only the time involved in running the actual dependency analysis on the maven projects. XtraLibD has an average response time of 12.043 seconds with a 61.37\% improvement in the response time over LibD while delivering higher response times than the rest of the tools.  

\subsection{Threats to validity}
\label{sec:threats-to-validity}
For developing our Lib2Vec models and $D$, we utilize a subset of Java TPLs (i.e., JAR files) present in the MavenCentral repository, and a subset of Python TPLs from Pypi.org. We assume that these TPLs cover a reasonably wide variety of Java and Python code such that the Lib2Vec models that we train on them will be accurate. However, there could still be many other Java and Python code types that could have improved the Lib2Vec models' accuracy. For training the Lib2Vec models, we obtain the normalized textual forms of the source code and bytecode representations of the .class files present in the JAR files and .py files present in Python TPLs (in .zip/ .rar/ /tar.gz forms). We obtain the source code and bytecode by using the compilation, decompilation and disassembly operations. Therefore, our Lib2Vec models' accuracy is subject to the accuracy of the decompilation and disassembly operations.

Next, we treat an unseen TPL that shows considerable similarity with the set of \enquote{available TPLs} as \emph{likely-to-be-non-irrelevant-TPLs}. Thus, the labeling of a TPL as \emph{likely-to-be-non-irrelevant-TPL} or \emph{irrelevant-TPL} is strongly dependent on its use in the considered application. We do not consider the TPL-usage as per now, but have included it as part of our future work. 

While training the Lib2Vec models, we consider only the .class and .py files of size 1kB or larger. However, there may exist Java and Python libraries where the average class size is lower than this limit. Excluding such a group of TPLs from the training might give inaccurate results when the input TPL being checked happens in such a group. The main reason for excluding such tiny source files is that they do not give sufficient Java and byte code, which is necessary to compute a sufficiently unique vector representation of the source code contents.

By reviewing the literature \cite{ma2016libradar,backes2016reliable,li2017libd}, we realized that there are a significant amount of TPL-detection tools designed for Android Applications, requiring the input file in an APK format. To the best of our knowledge, very few tools perform the TPL-detection for software applications existing in JAR format or for Python applications. Therefore, we converted our TPLs present from JAR  to APK format using the Android Studio toolkit and choose LibD, LibRadar, and LibScout -- some of the popular TPL-detection tools for our comparison. However, due to the fast advances of research in this area, we might have missed some interesting TPL-detection tool that works with the JAR file formats.

\section{CONCLUSIONS}
\label{sec:concl}
Software Development is immensely supported by the functionalities provided by various TPLs. However, as the software progresses through various releases, there tend to remain some irrelevant TPLs in the software distributable. These irrelevant TPLs lead to the unnecessary consumption of various resources such as CPU cycles, memory, etc., and thus its desirable to remove them. We propose a novel extra-library detector (XtraLibD) tool that detects the irrelevant libraries present in an application distributable. XtraLibD detects these libraries by performing code similarity detection with a reference dataset of 450 Java and 43,711 Python TPLs collected from MavenCentral. We use PVA to train language-specific code similarity detection model on the source code and byte code of these MavenCentral libraries. To the best of our knowledge, we are the first to apply PVA for detecting code similarity in TPLs. 

We used source code and byte code representations of TPLs to train our models as these preserve the semantics and are free from source code obfuscations. We successfully leveraged the semantics-preserving Java decompilers (and Python compilers) to transform the binary .class files and .pyc files into an obfuscation-invariant textual form, viz., source code and byte code. We verified our approach's efficacy by testing it with more than 30,000 .class files and 478,031 .py files, where we have achieved detection accuracy above 99.48\% and an F1 score of 0.968 for Java, and 99.5\% accuracy for Python. XtraLibD outperforms the existing TPL-detection tools, such as LibScout, LiteRadar, and LibD, with an accuracy improvement of 74.5\%, 30.33\%, and 14.1\%, respectively. XtraLibD achieves a storage reduction of 87.93\% over LibD and of 99.85\% over JIngredient. 

As part of the future work, we plan to explore the direction of actual TPL-usage within the application to detect the unused parts of TPL-code. The idea can be leveraged to develop various software artifacts for automating the SDLC activities, such as software code review, source code recommendation, and code clone detection.

%
% the environments 'definition', 'lemma', 'proposition', 'corollary',
% 'remark', and 'example' are defined in the LLNCS documentclass as well.
%

\bibliographystyle{splncs04}
\bibliography{bibFile}

\section*{APPENDIX}
\label{ap:parameter-tuning-PVA}
\textbf{Objective:}
The objective here is to seek an answer to our questions:
\begin{enumerate}
    \item \textit{For reliably detecting a library, what is the acceptable value of Lib2Vec similarity scores for source code and bytecode inputs?}
    \item \textit{Does the threshold similarity score ($\hat{\alpha}$) vary with the input parameters ($\beta, \gamma$, and $\psi$) of PVA?}
    \item \textit{What are the optimal values for the PVA tuning-parameters $\beta, \gamma,$ and $\psi$?} 
\end{enumerate}

Please refer to Table-\ref{tab:BloatLibD-TableOfNotation} for notation definitions.

\begin{table}[h]
    \centering
    %\scriptsize
    \caption{Scenarios for training Lib2Vec models using PVA \cite{dewan2021bloatlibd}.}
    \label{table:training-scenarios}
    \resizebox{0.7\columnwidth}{!}
    {
    \begin{tabular}{|c| c| c| c|  }
    \cline{1-3}
    \multicolumn{2}{c}{\textbf{Parameters varied}} & \multicolumn{1}{c}{} \\ \hline
    \textbf{Epochs $\beta$} & \textbf{Vector size $\gamma$}    & \textbf{Training samples $\psi$} & \textbf{Models}\\ \hline
    \multirow{3}{*}{\shortstack{Fixed\\ at\\ 10}} & \multirow{3}{*}{\shortstack{Fixed\\ at\\ 10}} & \multirow{3}{*}{\shortstack{Vary 5000 to-\\ \texttt{CorpusSize} in-\\ steps of 5000}} & \multirow{3}{*}{\shortstack{\texttt{CorpusSize}\\ $\div$ \\ 5000}} \\ 
     &&&\\
    &&&\\\hline
    \multirow{3}{*}{\shortstack{Vary 5-\\ to 50 in-\\ steps of 5}} & \multirow{3}{*}{\shortstack{Fixed\\ at\\ 10}} & \multirow{3}{*}{\shortstack{Fixed at-\\ \texttt{CorpusSize}}} & \multirow{3}{*}{10}\\ 
     &&&\\
     &&&\\
    \hline
    \multirow{3}{*}{\shortstack{Fixed\\ at\\ 10}} & \multirow{3}{*}{\shortstack{Vary 5-\\ to 50 in-\\ steps of 5}} & \multirow{3}{*}{\shortstack{Fixed at-\\ \texttt{CorpusSize}}}& \multirow{3}{*}{10} \\ 
      &&&\\
    &&&\\
    \hline
    \end{tabular}}
\end{table}

\begin{figure*}
\centering
\includegraphics[width=0.9\linewidth]{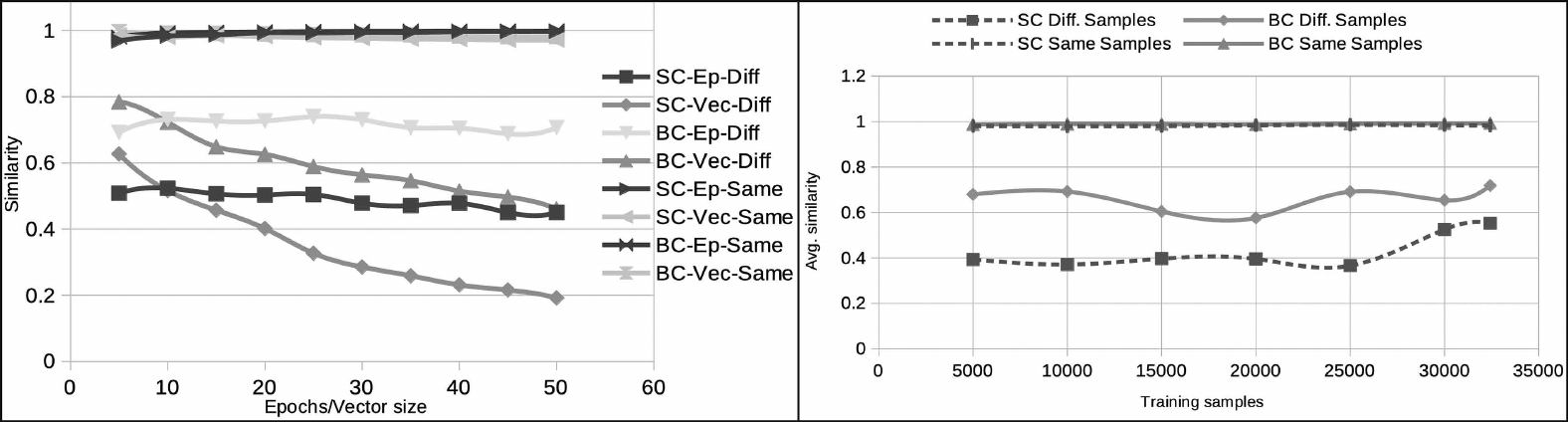} 

\caption{Variation of \emph{average similarity} with PVA tuning-parameters \cite{dewan2021bloatlibd}.}
\label{fig:Lib2Vec-similarity-variation}
\end{figure*}

\begin{figure*}
\centering
\includegraphics[width=0.9\linewidth]{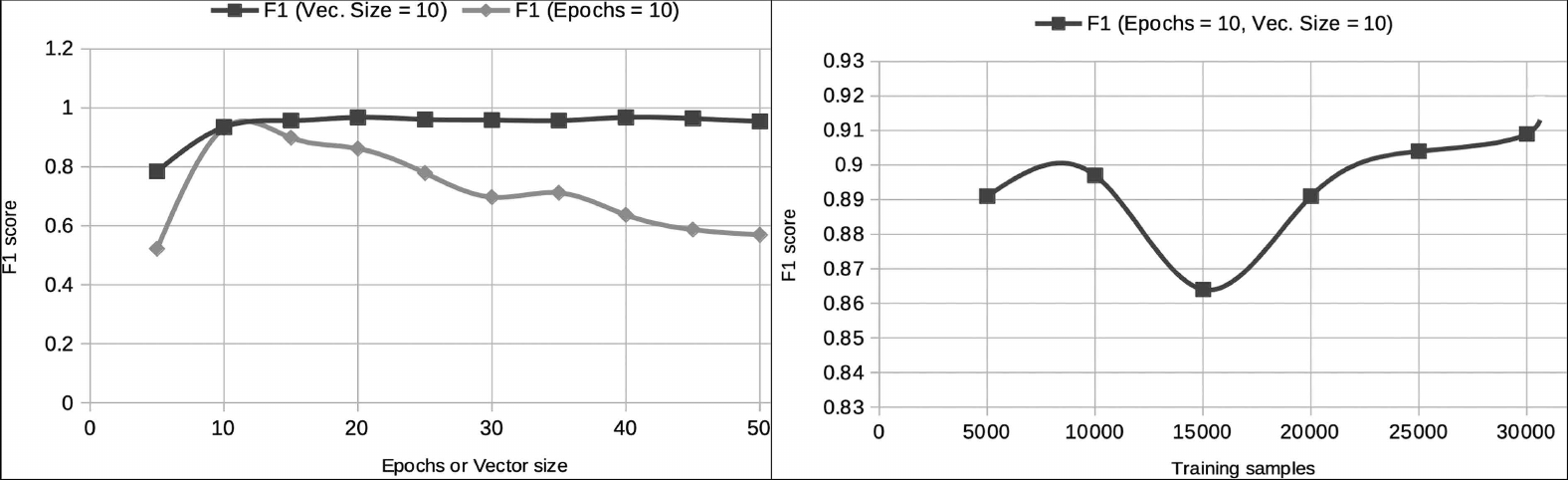} 

\caption{Performance metrics with PVA models trained on \emph{source code \cite{dewan2021bloatlibd}.}}
\label{fig:Lib2Vec-parameter-tuning}
\end{figure*}

\textbf{Test-bed setup:} Using the test partition of the test-bed developed in Section~\ref{sec:test-bed-setup}, we generate a test dataset ($Y$) containing $same, different$ file pairs in 50:50 ratio. Further, to check if our tool is resilient to source code transformations, we test it for the following three scenarios:
\begin{enumerate}
    \item \textit{Package-name transformations:} Package names of the classes present in TPLs are modified.
    \item \textit{Function (or method) transformations:}  Function names are changed in the constituent classes' source code, and function bodies are relocated within a class. 
    \item \textit{Source code transformations:} Names of various variables are changed, source code statements are inserted, deleted, or modified, such that it does not alter the semantics of the source code. For instance, adding print statements at various places in the source file.
\end{enumerate}
We test Lib2Vec models' efficacy in detecting similar source code pairs (or bytecode pairs) using $Y$.

\textbf{Procedure:}
%To compute the optimal values of $\beta, \gamma$ and $\psi$ and the threshold ($\hat{\alpha}$) of Lib2Vec similarity scores, we perform the following steps:
The salient steps are \cite{dewan2021bloatlibd}:
\begin{enumerate}
\item $F_{bc}, F_{sc} := $ Obtain the textual forms of bytecode and source code present in source files of training JARs of the test-bed (developed in Section~\ref{sec:test-bed-setup}).
\item For each parameter combination $\pi \in Z$ (listed in Table-\ref{table:training-scenarios}):
\begin{enumerate}
    \item  $S_{sc}^{\pi} := S_{bc}^{\pi} := NULL$
    \item $M_{bc}^{\pi}, M_{sc}^{\pi} := $ Train the Lib2Vec models using $F_{bc}, F_{sc}$.
    \item  Save $M_{bc}^{\pi}$ and $M_{sc}^{\pi}$ to disk.
    \item For each file pairs $\langle p_i, p_j \rangle \in Y$:
    \begin{enumerate}
        \item $\phi_{i}, \phi_{j} := $ Obtain PVA vectors for $p_i,p_j$ using $M(\pi)$
        \item $\alpha_{i,j} :=$ Compute cosine similarity between  $\langle \phi_{i}, \phi_{j} \rangle$
        \item if $p_i == p_j$: $S_{same} = S_{same} \cup \langle \alpha_{i,j}  \rangle$
        \item else: $S_{different} = S_{different} \cup \langle \alpha_{i,j}  \rangle$
    \end{enumerate}
    \item $\hat{\alpha_{bc}^{\pi}}, \hat{\alpha_{sc}^{\pi}} := $ Obtain the average similarity scores using  $S_{bc}^{\pi}$ and $S_{sc}^{\pi}$ and save them.
    \item Using the $\hat{\alpha_{bc}^{\pi}}, \hat{\alpha_{sc}^{\pi}}$ as thresholds, compute the accuracy of  $M_{bc}^{\pi}$ and $M_{sc}^{\pi}$.
    \item Plot the variation of $\hat{\alpha_{bc}}$, $\hat{\alpha_{sc}}$, the accuracy of PVA models for different values of $\beta, \gamma,$ and $\psi$ used in the experiment, and analyze.
\end{enumerate}
\end{enumerate}

\textbf{Results and Observations:}
%Due to the limitation of space, we have shared our result plots at \url{https://bit.ly/3mGCess}. 
Figure \ref{fig:Lib2Vec-similarity-variation} and \ref{fig:Lib2Vec-parameter-tuning} show the effect of PVA tuning-parameters on the \emph{average similarity} and the model performance metrics values, respectively. The legend entry BC-Ep-Diff represents the similarity variation w.r.t epochs for bytecode case when two samples were different. SC-Vec-Same indicates the variation w.r.t vector size for source code case when two samples were identical. The following are the salient observations:
\begin{enumerate}
    %\item The similarity between the vectors tends to decrease as we increase $\gamma$ above 5, and $\beta$ above 10. 
    \item Effect of increasing the epochs beyond 10 seems to have a diminishing improvement in the accuracy scores.
    \item A noticeable decrease in similarity scores was observed by increasing the vector count beyond 5, and the epochs count beyond 10.
    \item As anticipated, the accuracy (indicated by F1 scores\footnote{\url{https://bit.ly/3kHqkNg}}) improves with the size of training samples.
\end{enumerate}

Therefore, we take $\hat{\alpha_{sc}}=0.98359$ and $\hat{\alpha_{bc}}=0.99110$ as the similarity threshold values for source code data and bytecode data, respectively. Further, the best accuracy (99.48\% for source code and 99.41\% for bytecode) is achieved with the Lib2Vec model trained using 30427 samples, 10 epochs, and the vector size of 10. The precision and recall values, in this case, were 99.00\% and 99.00\%, respectively, resulting in an F1 score of 99\% for the source code case. 
As we achieve the highest accuracy scores at $\beta=\gamma=10$, we take these as the optimal input parameter values for PVA.

\end{document}